\documentclass[a4paper]{jpconf}
\usepackage{amsmath,natbib,blackboard,subfig}
  \usepackage[pdftex]{graphicx}
  \graphicspath{{figures/png/},{figures/jpg/}}
  \DeclareGraphicsExtensions{.pdf,.jpg,.png}
\def\vec#1{{\bf #1}}
\def\op#1{\hat{#1}}
\def\ket#1{| #1 \rangle}
\def\bra#1{\langle #1 |}

\def\ave#1{\langle #1 \rangle}

\font\liouvop=eufm10

\def\SU{{\text{\liouvop SU}}}
\def\sl{{\text{\liouvop sl}}}

\def\D{\mathcal{D}}

\def\H{\mathcal{H}}
\def\L{\mathcal{L}}

\def\eps{\epsilon}
\def\sx{\op{\sigma}_x}
\def\sy{\op{\sigma}_y}
\def\sz{\op{\sigma}_z}
\def\s{\op{\sigma}}

\def\norm#1{\| #1 \|}

\begin{document}
\title{Physics-based mathematical models for quantum devices via
experimental system identification}
\author{Sonia G Schirmer}
\address{Dept of Applied Maths \& Theoretical Physics,
University of Cambridge, \\ Wilberforce Rd, Cambridge, CB3 0WA, UK}
\ead{sgs29@cam.ac.uk}
\author{Daniel K L Oi}
\address{SUPA, Dept of Physics, University of Strathclyde, Glasgow G4 0NG, UK}
\author{Simon J Devitt}
\address{National Institute of Informatics, 2-1-2 Hitotsubashi, 
 Chiyoda-ku, Tokyo 101-8430, Japan}

\begin{abstract}
We consider the task of intrinsic control system identification for
quantum devices.  The problem of experimental determination of subspace 
confinement is considered, and simple general strategies for full
Hamiltonian identification and decoherence characterization of a 
controlled two-level system are presented.
\end{abstract}

\section{Introduction}

Advances in nano-fabrication are increasingly enabling us to create
nano-scale devices that exhibit non-classical or quantum-mechanical
behaviour.  Such quantum devices are of great interest as they may
pave the way for a new generation of quantum technology with various
applications from quantum metrology to quantum information processing.  
However, to create quantum devices that perform useful functions, we 
must be able to understand their behaviour, and have effective means 
to controllably manipulate it.  Analysis of system dynamics and the 
design of effective control strategies is almost impossible without 
the availability of sufficiently accurate mathematical models of the 
device.  While these models should capture the essential features of 
the device, to be useful, they must also be computationally tractable, 
and preferably as simple as possible.

There are different approaches to deal with this problem.  One, which we
shall refer to as a first-principles approach, involves constructing
model Hamiltonians based on reasonable assumptions about the relevant
physical processes governing the behaviour of the system, making various
simplifications, and solving the Schrodinger equation in some form,
usually using numerical techniques such as finite element or functional
expansion methods.  The empirical approach, on the other hand, starts
with experimental data and observations to construct a model of the
system.  In practice both approaches are needed to deal with complex
systems.  First-principle models are crucial to elucidate the
fundamental physics that governs a system or device, but experimental
data is crucial to account for the many unknowns that result in
variability of the characteristics of the system, which is often
difficult to predict or explain theoretically.  For instance,
`manufactured' systems such as artificial quantum dot `atoms' and
`molecules' vary considerably in size, geometry, internal structure,
etc, and sometimes even small variations can significantly alter their
behaviour.

\begin{figure}
\centering\includegraphics{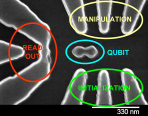}
\caption{Simple charge `qubit' device: Trench-isolated Silicon double 
quantum dot molecule fabricated using e-beam lithography, controlled by 
several DC-gate electrodes, with a single electron transistor for charge 
readout (Hitachi Cambridge Labs)~\cite{PhysRevLett.95.090502}.}
\label{fig:charge-qubit}
\end{figure}

In this paper we will focus on constructing mathematical models solely
from experimental data without any prior knowledge of the specifics of
the system, an approach one might call black-box system identification. 
We will restrict ourselves to quantum systems whose essential features
can be captured by low-dimensional Hilbert space models.  For instance, 
a quantum dot molecule such as the Silicon double quantum dot system in 
Fig.~\ref{fig:charge-qubit} may consist of millions of individual atoms 
with many degrees of freedom, but the observable dynamics may be well
described by overall charge distribution states in the quantum dot 
molecule.  Taking these charge distribution states as basis states for
a Hilbert space $\H$, the observable dynamics of the system is governed
by a Hamiltonian operator acting on $\H$, and possibly additional
non-Hermitian operators to account for dissipative effects if the system 
is not completely isolated from its environment.  If the dimension of
the Hilbert space $\H$ is huge then complete characterization of the
Hamiltonian and dissipation operators may be a hopeless task, but in 
certain cases the dynamics of interest takes place in low-dimensional 
Hilbert space $\H$, or we may in fact wish to \emph{design} a device 
whose dynamics can be described by a low-dimensional Hilbert space model, 
as is the case in quantum information processing~\cite{00Nielsen}, where 
we desire to create systems that act as quantum bits, for example.

The paper is organized as follows.  We start with a general description
of the basic assumptions we make about the system to be modelled and the
type of measurements and experiments upon which our characterization
protocols are based.  In section~\ref{sec:confinement} we discuss the
problem of subspace confinement, i.e., how to experimentally characterize 
how well the dynamics of system is confined to a low-dimensional subspace 
such as a two-dimensional (qubit) Hilbert space. In 
section~\ref{sec:qubit_ident} we present general protocols for
identifying Hamiltonian and decoherence parameters for controlled
qubit-like systems.  We conclude with a discussion of generalizations
and open problems.

\section{General Modelling Assumptions}
\label{sec:model}

We start with a generic model assuming only that the state of the system
can be represented by a density operator (i.e., a positive operator of 
trace one) $\op{\rho}$ acting on a Hilbert space $\H$ of dimension $N$, 
and that its evolution is governed by the quantum Liouville equation
(in units such that $\hbar=1$):
\begin{equation}
  \frac{d}{dt}\op{\rho}(t) 
  = -\rmi [\op{H},\op{\rho}(t)] + \L_D[\op{\rho}(t)],
\end{equation}
where $\op{H}$ is an effective Hamiltonian and $\L_D$ a super-operator 
that accounts for dissipative effects due to environmental influences, etc.
Initially, all that is known about $\op{H}$ and $\L_D$ is that $\op{H}$ 
is a Hermitian operator on $\H$, and $\L_D$ a completely positive 
super-operator acting on density matrices $\op{\rho}(t)$, although we 
may make some additional assumptions about the structure of $\op{H}$ or 
$\L_D$.  For instance, we shall generally assumes $\L_D$ to be of 
Kossakowski-Sudarshan-Lindblad form~\cite{76Lindblad,76Gorini}:
\begin{equation}
  \label{eq:diss}
  \L_D[\op{\rho}(t)] = \sum_{k=1}^{N^2-1} \gamma_k \D[\op{V}_k] \op{\rho}(t)
\end{equation}
where the super-operators $\D[\op{V}_k]$ are defined by
\begin{equation}
  \D[A]B = A B A^\dagger - (A^\dagger A B + B A^\dagger A)/2,
\end{equation}
$\op{V}_k$ are (generally non-Hermitian) operators on $\H$ and $\gamma_k$ 
are positive real numbers.

If the system is subject to dynamic control, then the Hamiltonian (and
sometimes the relaxation operators $\op{V}_k$) are not fixed but dependent 
on external control fields, which we shall denote by $\vec{f}(t)$.  In the
simplest case, we may assume a linear dependence of the Hamiltonian on the 
controls
\begin{equation}
   \label{eq:Hf}
   \op{H}[\vec{f}(t)] = \op{H}_0 + \sum_{m=1}^M f_m(t) \op{H}_m,  
\end{equation}
where the $\op{H}_m$ are fixed Hermitian operators on $\H$ and the 
reservoir operators $\op{V}_k$ are constant.   

The ultimate objective of characterization is to identify the operators
$\op{H}[\vec{f}(t)]$ (or $\op{H}_m$) and $\L_D$ by performing suitable
experiments on the system.  This task would be simplified if we could
assume that the system can be prepared in an arbitrary initial state
$\op{\rho}_0$, and that we can perform generalized measurements or
projective measurements in arbitrary bases at any time---but these
requirements are generally unrealistic.  For example, our ability to
perform measurements on the system prior to characterization is limited
by the direct readout processes available, and often we only have a
single, basic sensor such as a single electron transistor (SET)~\cite{NAT406p1039} 
providing relatively limited information about the charge distribution 
in a quantum dot molecule, or detection may be accomplished via a readout
transition that involves coupling one state of the system to a
fluorescent read-out state, e.g., via a laser, etc.  Furthermore,
preparation of non-trivial states generally depends on knowledge of the
operators $\op{H}[\vec{f}(t)]$ and $\L_D$, the very information about
the system we are trying to obtain.  These practical restrictions rule
out conventional quantum state or process tomography techniques, which
presume the ability to measure the system in different measurement bases
and prepare it in different initial states to obtain sufficient
information to reconstruct the quantum state or 
process~\cite{JMO44p2455,PRL78p0390,JMP44p0528,PhysRevA.66.012303,qph0411093}.

In this paper we consider a rather typical experimental scenario, where
we are limited to measurements of a fixed observable and evolution under 
a Hamiltonian that can be modified by varying certain control settings. 
For the most part, we restrict ourselves further to piecewise constant 
controls.  The only assumption on the measurement process we make is that
it can be formally represented by some Hermitian operator $\op{A}$ with 
$N$ eigenvalues corresponding to measurement outcomes $\lambda_n$, i.e., 
that it has a spectral decomposition of the form
\begin{equation}
 \label{eq:A}
   \op{A} = \sum_{n=1}^N \lambda_n \ket{n}\bra{n}.
\end{equation}
This measurement also serves as initialization of the system as outcome 
$\lambda_n$ means that the system will be left in an eigenstate $\ket{n}$ 
associated with the eigenvalue $\lambda_n$.  If $\op{A}$ has $N$ unique 
eigenvalues, i.e., all eigenvalues occur with multiplicity one, then the 
measurement is sufficient to initialize the system in a unique state; if
it has degenerate eigenvalues associated with eigenspaces of dimension 
greater $>1$ then some measurement outcomes will not determine the state 
uniquely.  

Following the idea of intrinsic characterization, our objective is to
extract information about the system without recourse to any external
resources, i.e., using no information from measurements other than the
information provided by the sensors built into the device, and no
external control fields except the ability to change the settings of the
built-in actuators (such as variable gate voltages) subject to
constraints.  Although this requirement of characterization relying only
on the built-in sensors and actuators may seem excessively restrictive,
excluding many forms of spectroscopy, for example, it has the advantage
of simplicity (no external resources required).  Moreover, external
sensors and actuators may disturb the system, and thus characterization
of the system in their presence may not yield an accurate picture of the
dynamics in the absence of the additional apparatus.

We restrict ourselves here to systems sufficiently weakly coupled to a
sufficiently large reservoir, whose dynamics can be described by adding
a dissipation super-operator of the form~(\ref{eq:diss}) to the
Hamiltonian dynamics of the subsystem of interest.  Although many
systems can be modelled this way, it should be noted that this approach
has limitations.  For example, the dynamics of a subsystem $\H_S$ that
is strongly is coupled to a finite reservoir $\H_R$, such as single spin
coupled to several nearby spins, can be very complicated and
non-Markovian.  Although non-Markovian dynamics can in principle be
dealt with by allowing time-dependent relaxation operators $\op{V}_k$,
in such cases it is not always possible to describe the dynamics of the
subsystem of interest in terms of the Hamiltonian dynamics on the
subspace and a set of simple relaxation or decoherence operators.
Rather, it may become necessary to consider the system $\H_S+\H_R$ and
characterize its dynamics, described by a Schrodinger equation with a
Hamiltonian $\op{H}_{S+R}$, instead to obtain an accurate picture of the
subsystem dynamics.  The basic ideas of intrinsic characterization can
be applied to this larger system, and the protocols we will describe can
in principle be extended to such higher-dimensional systems, although
full characterization of the system plus reservoir Hamiltonian
$\op{H}_{S+R}$ using only the built-in sensors and actuators may not be
possible.  The degree of characterization possible will depend on the
size of the Hilbert space and the capabilities of the built-in sensors
and actuators, e.g., to discriminate and manipulate different states of
the larger system.  Before attempting to identify the system Hamiltonian
(and decoherence operators), an important first task is therefore
estimating the dimension of the Hilbert space in which the dynamics
takes place.  

\section{Characterization of subspace confinement}
\label{sec:confinement}

A fundamental prerequisite for constructing a Hilbert space model is 
knowledge of the underlying Hilbert space.  This is a nontrivial problem
as most systems have many degrees of freedom, and thus a potentially 
huge Hilbert space, but effective characterization of the system often
depends on finding a low dimensional Hilbert space model that captures 
the essential features of the system.  Furthermore, in applications such 
as quantum information processing the elementary building blocks are 
required to have a certain Hilbert space dimension.  For example, for a
quantum-dot molecule to qualify as a qubit, we must be able to isolate a
two-dimensional subspace of the total Hilbert space, and be able to
coherently manipulate states within this 2D subspace without coupling to 
states outside the subspace (leakage).  This requires several 
characterization steps:
\begin{enumerate}
\item Isolation of a 2D subspace and characterization of subspace 
      confinement;
\item Characterization of the Hamiltonian dynamics including effect
      of the actuators; and
\item Characterization of (non-controllable) environmental effects 
      (dissipation).
\end{enumerate}

The choice of a suitable subspace depends also on the measurement device
as the measurement must be able to distinguish the basis states.  Thus
for a potential charge qubit device, for example, only a subspace
spanned by charge states that can be reliably distinguished by the SET
is a suitable candidate, and ideally there should be two (or in general
$N$) orthogonal states that can be perfectly distinguished by the
measurement, so that we have a well-defined reference frame (basis)
$\{\ket{0},\ket{1},\ldots,\ket{N}\}$, and the measurement can be
represented by an observable of the form~(\ref{eq:A}) with $N$ distinct
eigenvalues $\lambda_n$.  Furthermore, we would like the readout process
to act as a projective measurement, so that each measurement projects
the state of the system onto one of the measurement basis states
$\ket{n}$.  The ability to perform projective measurements is not a
trivial requirement, especially for solid state systems, and is not an
absolute necessity as characterization protocols can be adapted to weak
measurements, but projective measurements can in principle be achieved
with various sensors such as rf-SETs, and we shall assume here that we
are working in this regime.

After a possible subspace has been identified, it is crucial to check
that the subspace is sufficiently isolated, i.e., that we can reliably 
initialize the system in a state in this subspace and that it remains 
in this subspace under both free and controlled evolution.  This 
characterization of subspace confinement is very important.  If the 
device is a candidate for a qubit, for example, it is essential that 
subspace leakage be much less than the rate of bit or phase flip errors 
due to imperfect control or decoherence as leakage or loss error 
correction protocols are considerably more demanding in terms of 
complexity and resources required.

\begin{figure}
\centering{\includegraphics{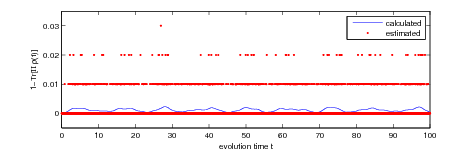}}
\caption{Leakage characterization assuming three-outcome measurement for
 system initialized in state $\ket{0}$.  The continuous curve is the
calculated value of $p_{\rm leak}(t)=1-\Tr[\Pi \op{\rho}(t)]$, where 
$\Pi=\ket{0}\bra{0}+\ket{1}\bra{1}$ is the projector onto the target
subspace, for a 10-level test Hamiltonian.
The red dots are the estimates of $p_{\rm leak}(t)$ at the discrete
times $t_k=(k-1)\Delta t$ with $\Delta t=0.01$, obtained by averaging 
over $N_e'=100$ simulated experiments for each $t_k$.  As the figure
shows, the effect of the projection noise makes it difficult to obtain 
accurate estimates of $p_{\rm leak}(t)$ for any given $t=t_k$ unless 
$N_e'$ is extremely large.  However, the average leakage rate over the 
time interval $[0,100]$, $\bar{p}_{\rm leak}=0.11$\% for both the 
calculated and measured values.}
\label{fig:leakage1}
\end{figure}

Characterization of subspace confinement depends on the characteristics
of the sensors, i.e., the type of measurements we can perform.  For our
charge qubit with SET readout example, if the SET can be calibrated to
be sufficiently sensitive to enable detection of states outside the 
chosen subspace in addition to being able to discriminate the subspace 
basis states $\ket{n}$, i.e., if the true measurement has (at least) 
$N+1$ mutually exclusive outcomes $\lambda_n$, $n=1,\ldots,N$ and 
$\lambda_{N+1}$ if the system state is outside the subspace, then the
characterization of subspace confinement is relatively easy.  If we 
perform $N_e$ experiments of the form
\begin{enumerate}
\item \textbf{Initialize:} Measure and record outcome $\lambda_a$ 
\item \textbf{Evolve:} Let the system evolve for time $t$ under some
      fixed Hamiltonian $H_{\vec{f}}$
\item \textbf{Measure:} Repeat measurement and record outcome
       $\lambda_b$
\end{enumerate}
and denote by $N_{a,b}$ the number of times the measurement results for 
the initial and final measurements were $\lambda_a$ and $\lambda_b$, 
respectively, then, for $N_e$ sufficiently large, the leakage out of the 
subspace is approximately equal to the fraction of experiments for which 
the first measurement was $\lambda_n$ with $n\le N$ and the second 
measurement was $N+1$, i.e., 
\begin{equation}
  p_{\rm leak} \approx \frac{\sum_{n=1}^N N_{n,N+1}}{N_e -\sum_{n=1}^{N+1} N_{N+1,n}}.
\end{equation}
Repeating the experiments for different evolution times $t$ and
averaging then gives an indication of the rate of subspace leakage,
e.g., if we estimate $p_{\rm leak}(t_k)$ for $t_k=k\Delta t$, 
$k=0,\ldots,K$, then
\begin{equation}
  \bar{p}_{\rm leak} = 
   \frac{1}{t_{\rm max}} \int_0^{t_{max}} p_{\rm leak}(t) \, dt
   \approx \frac{1}{K+1} \sum_{k=0}^K p_{\rm leak}(t_k). 
\end{equation}

\begin{figure}
\centering{\includegraphics{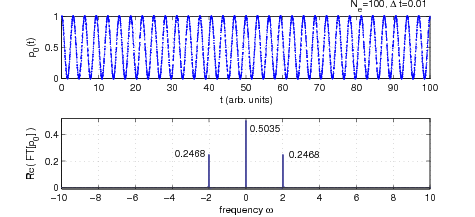}}
\caption{$p_0(t)$ obtained by averaging over $N_e=100$ simulated 
experiments with a 10-level test Hamiltonian and corresponding 
Fourier spectrum.  From the 0th and 1st order peak heights we obtain
$h_0+2h_1=0.9970$, which yields upper and lower bounds for $\eps$ of
$\approx 0.0015$, although these bounds are statistically uncertain 
due to projection noise and discretization errors.  Nonetheless, it 
can be verified that the actual subspace confinement for the chosen
test Hamiltonian is $99.94$\%, and thus the upper bound provides a 
reasonable estimate for the leakage.}
\label{fig:leakage2}
\end{figure}

If the SET (or other measurement process) is not sufficiently sensitive
to reliably distinguish at least $N$ subspace basis states $\ket{n}$ as
well as states outside the subspace, then subspace characterization is 
more challenging.  For instance, suppose we have a 2D subspace and a sensor 
that can reliably detect only one state, say $\ket{0}$, a common scenario
for many systems where readout transitions are used that detect only a
single state.  In this case the measurement outcomes are $\lambda_0$ or 
$\neg \lambda_0$.  If the measurement is projective and the dynamics
confined to a 2D subspace, we can identify $\neg\lambda_0$ with outcome 
$\lambda_1$ but this identification will lead to errors if the dynamics
is not really confined to a 2D subspace.  However, even in this case we 
can still estimate the level of confinement to a 2D subspace under the
evolution of a Hamiltonian $\op{H}_{\vec{f}}$ from observable coherent
oscillations.  For example, let
\begin{equation}
  p_0(t) = |\bra{0} e^{-i\op{H}_{\vec{f}}t}\ket{0}|^2
\end{equation}
be the probability of obtaining outcome $\lambda_0$ when measuring the 
time-evolved state $\ket{\Psi(t)}=\exp(-i t\op{H}_{\vec{f}})\ket{0}$.
If the system is initialized in the state $\ket{0}$ and the dynamics of 
the system under $\op{H}_{\vec{f}}$ is perfectly confined to a two-level
subspace then conservation of probability implies that the heights $h_0$ 
and $h_1$ of the 0th and 1st order terms in the Fourier spectrum of 
$p_0(t)$ must satisfy $h_0+2h_1=1$.  We can use the deviation from this
equality to bound the subspace leakage $\eps$~\cite{NJP7p384}:
\begin{equation}
  1-\sqrt{h_0+2h_1} \le \eps \le \frac{1}{2}(1-\sqrt{2(h_0+2h_1)-1}).
\end{equation}
As the upper and lower bounds depend only on the 0th and 1st order 
Fourier peaks, they can usually be easily determined from experimental
data although finite resolution due to discretization and projection 
noise will reduce the accuracy of the estimates~\cite{NJP7p384}.
Moreover, in this case $1-\eps$ only indicates the confinement of the 
system to \emph{some} two-level subspace, and we must be careful as 
for different controls $\vec{f}$ the dynamics under $\op{H}_{\vec{f}}$ 
may be confined to \emph{different} two-level subspaces, in which case 
the system cannot be considered a proper qubit.  For instance, for the
10-level test Hamiltonian 
\begin{equation*}
  \op{H}_{\rm test} = \begin{pmatrix}
  1.3701 & 1.0000 & 0.0093 & 0.0055 & 0.0112 & 0.0068 & 0.0119 & 0.0084 & 0.0065 & 0.0087\\
  1.0000 & 1.5561 & 0.0109 & 0.0132 & 0.0067 & 0.0061 & 0.0081 & 0.0051 & 0.0105 & 0.0029\\
  0.0093 & 0.0109 & 1.6603 & 0.0034 & 0.0161 & 0.0100 & 0.0101 & 0.0123 & 0.0115 & 0.0055\\
  0.0055 & 0.0132 & 0.0034 & 1.9112 & 0.0136 & 0.0072 & 0.0093 & 0.0062 & 0.0133 & 0.0101\\
  0.0112 & 0.0067 & 0.0161 & 0.0136 & 3.4611 & 0.0022 & 0.0119 & 0.0078 & 0.0064 & 0.0122\\
  0.0068 & 0.0061 & 0.0100 & 0.0072 & 0.0022 & 4.3017 & 0.0074 & 0.0077 & 0.0029 & 0.0080\\
  0.0119 & 0.0081 & 0.0101 & 0.0093 & 0.0119 & 0.0074 & 6.8732 & 0.0133 & 0.0158 & 0.0154\\
  0.0084 & 0.0051 & 0.0123 & 0.0062 & 0.0078 & 0.0077 & 0.0133 & 7.3491 & 0.0071 & 0.0073\\
  0.0065 & 0.0105 & 0.0115 & 0.0133 & 0.0064 & 0.0029 & 0.0158 & 0.0071 & 8.1876 & 0.0108\\
  0.0087 & 0.0029 & 0.0055 & 0.0101 & 0.0122 & 0.0080 & 0.0154 & 0.0073 & 0.0108 & 8.9032
\end{pmatrix}
\end{equation*}
used for the simulated experiments in Figs~\ref{fig:leakage1} and
\ref{fig:leakage2}, the first procedure measures the projection onto 
the subspace $S_1$ spanned by the measurement basis states
$\ket{0}=(1,0,0,0,0,0,0,0,0,0,0)^T$ and $\ket{1} =
(0,1,0,0,0,0,0,0,0,0,0)^T$,  while the second procedure measures the
confinement of the dynamics to the best-fitting 2D subspace $S_2$, which
for the given Hamiltonian is spanned by
\begin{align*}
 \vec{v}_1 &= (1,-0.0001, 0.0004, 0.0034,-0.0012,-0.0002,-0.0005,-0.0004, 0.0003,-0.0005)^T,\\ 
 \vec{v}_2 &= (0, 1, 0.0174, 0.0238,-0.0131,-0.0051,-0.0032,-0.0020,-0.0022,-0.0013)^T,
\end{align*}
and comparison of the results confirms that the average confinement 
$1-\bar{p}_{\rm leak}$ for the subspace $S_1$ as obtained by the first 
procedure is less ($99.89$\%) than the confinement for subspace $S_2$ 
($99.94$\%).

\section{General characterization protocols for qubit systems}
\label{sec:qubit_ident}

Once a suitable subspace has been chosen, we can proceed to the second
stage of the characterization, identification of the Hamiltonian and
decoherence operators.  The simplest type of system we can consider here
is a qubit with Hilbert space dimension $2$.  In this case, any
non-trivial projective measurement, i.e., any measurement with two
distinguishable outcomes $\lambda_0$ and $\lambda_1$ can be represented
by an observable
$\op{A}=\lambda_0\ket{0}\bra{0}+\lambda_1\ket{1}\bra{1}$.  The
eigenstates $\ket{0}$ and $\ket{1}$ of $\op{A}$ define a basis for the
Hilbert space, and we can define the Pauli operators $\s_0$, $\sx$,
$\sy$ and $\sz$ with respect to this basis
\begin{subequations}
\begin{align}
  \s_0 = \ket{0}\bra{0} + \ket{1}\bra{1}, & \quad 
  \sz  = \ket{0}\bra{0}-\ket{1}\bra{1},\\
  \sx  = \ket{0}\bra{1} + \ket{1}\bra{0}, & \quad 
  \sy  = \rmi(-\ket{0}\bra{1}+\ket{1}\bra{0}),
\end{align}
\end{subequations}
or in the usual matrix notation
\begin{equation}
   \s_0= \begin{pmatrix} 1 & 0 \\ 0 & 1 \end{pmatrix}, \;
   \sx = \begin{pmatrix} 0 & 1 \\ 1 & 0 \end{pmatrix}, \;
   \sy = \begin{pmatrix} 0 & -\rmi \\ \rmi & 0 \end{pmatrix}, \;
   \sz = \begin{pmatrix} 1 & 0 \\ 0 & -1 \end{pmatrix}.
\end{equation}
Taking (without loss of generality) the measurement outcomes to be 
$\lambda_0=1$ and $\lambda_1=-1$, we have $\op{A}=\sz$ in this basis.  
Furthermore, we can expand any Hamiltonian of the system as
\begin{subequations} 
\label{eq:H2}
\begin{align}
  \op{H} &= \frac{1}{2} (d_0 \s_0 + d_x \sx + d_y \sy + d_z \sz) \\
         &= \frac{d_0}{2} \s_0 + \frac{\omega}{2} 
            (\sin\theta\cos\phi\,\sx + \sin\theta\sin\phi\,\sy + \cos\theta\,\sz).
\end{align}
\end{subequations}
The $d_0$ term can generally be ignored as the identity $\s_0$ commutes
with all other Pauli matrices and $\exp(-\rmi t d_0 \s_0)$ corresponds 
to multiplication by a global phase factor, i.e., trivial dynamics.
Hence, it suffices to determine the real vector $\vec{d}=(d_x,d_y,d_z)$,
or in polar form, the angles $(\omega,\theta,\phi)$ to determine the
Hamiltonian.  For a single Hamiltonian, we can furthermore choose the
coordinate system such that $\phi=0$.  However, since $\op{H}$ depends
on control inputs $\vec{f}$, the parameters $\omega$, $\theta$ and 
$\phi$ also depend on the controls $\vec{f}$, and we usually need to
determine $\op{H}_{\vec{f}}$ for many different $\vec{f}$.  In this case
we can choose a reference Hamiltonian, e.g., $\op{H}_{\rm ref}$, with 
$\phi=0$, but we must determine the relative angles $\phi$ with respect 
to the reference Hamiltonian for all other control settings.

\subsection{Rotation frequency and declination of rotation axis}

\begin{figure}[!t]
\centerline{%
\subfloat[]{\includegraphics[height=2.2in]{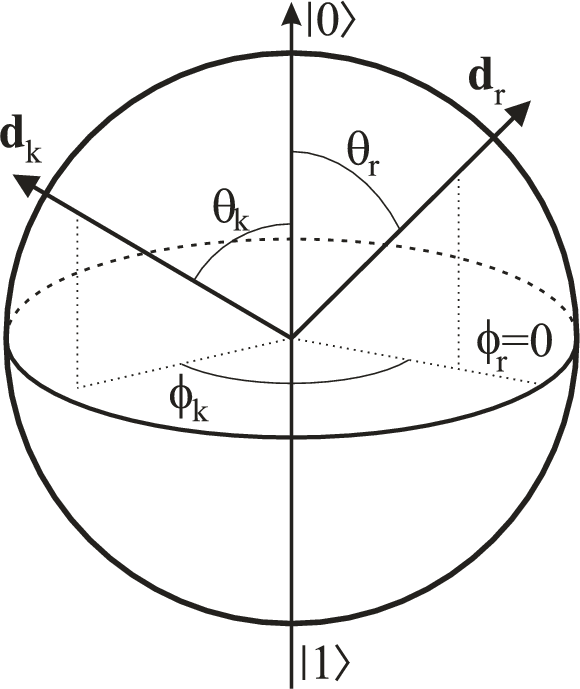} 
\label{fig_first_case}} \hfil
\subfloat[]{\includegraphics[height=2.2in]{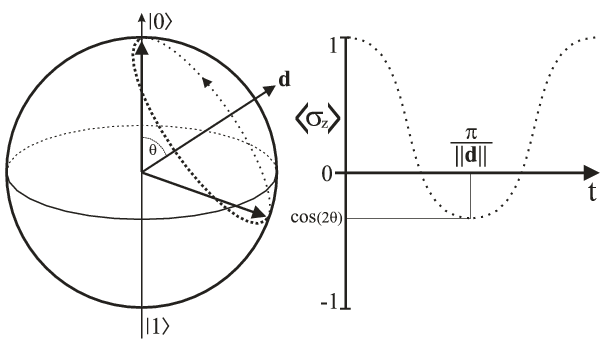}
\label{fig_second_case}}}
\caption{Bloch sphere representation of dynamics.  (a) Arrangement
of rotation axes $\vec{d}_k$ with respect to reference rotation axis
$\vec{d}_r$.  (b) The evolution of the vector $\vec{s}(t)$ with 
$\vec{s}_0=(0,0,1)^T$ about the axis $\vec{d}$, and the projection 
of $\vec{s}(t)$ onto the $\vec{z}$-axis.}
\label{fig:bloch}
\end{figure}

\begin{figure}
\center\scalebox{0.5}{\includegraphics{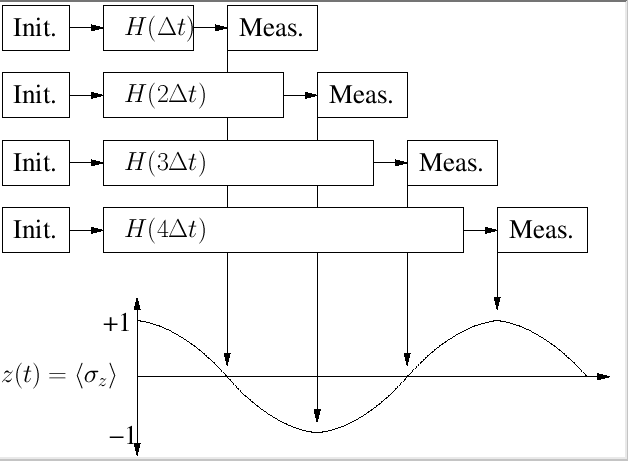}}
\caption{Stroboscopic mapping of coherent oscillations with projective 
measurements.  After initialization through measurement, the system is 
allowed to evolve for fixed times under the influence of the controls 
before a second measurement is taken.  The experiments are repeated to 
determine the expectation value of the observable at each time.}
\label{fig:stobo}
\end{figure}

To relate the Hamiltonian parameters to observable dynamics, it is
instructive to visualize qubit states and their evolution on the Bloch 
sphere.  If we define $\vec{s}=(s_x,s_y,s_z)^T$ with $s_k=\Tr(\op{\rho}\s_k)$ 
for $k\in\{x,y,z\}$ then it is easy to check that there is a one-to-one 
correspondence between density operators $\op{\rho}$ and points $\vec{s}$ 
inside the closed unit Ball in $\RR^3$.  Furthermore, the evolution of 
$\rho(t)$ under the (constant) Hamiltonian~(\ref{eq:H2}) corresponds to 
a rotation of $\vec{s}(t)$ about the (unit) axis 
$\hat{\vec{d}}=(\sin\theta\cos\phi,\sin\theta\sin\phi,\cos\theta)^T$ with 
angular velocity $\omega$ as illustrated in Fig.~\ref{fig:bloch}.  The
figure also shows that we can determine the angle $\theta$ and angular 
frequency $\norm{\vec{d}}=\omega$ from the projection of $\vec{s}(t)$ 
under the rotation about the axis $\vec{d}$ onto the $\vec{z}$-axis.
As the figure shows, given $z(t)$, we can in principle extract $\omega$ 
and $\theta$ from the first minimum $(t_0,z_0)$ of $z(t)$, although in 
practice it is usually preferable to use Fourier analysis or 
harmonic inversion techniques.
$z(t)$, being the expectation value $\ave{A(t)}=\Tr[\sz\rho(t)]$ of the 
observable $\op{A}=\sz$ in the state $\op{\rho}(t)$, can be obtained 
experimentally as follows~\cite{PRA69n050306}: 
\begin{enumerate}
\item \textbf{Initialize:} Measure and record outcome $\lambda_a=\pm 1$ 
      $\Rightarrow$ system in state $\lambda_a\vec{s}_0=\lambda_a(0,0,1)^T$.
\item \textbf{Evolve:} Let the system evolve for time $t$ under the 
      control settings $\vec{f}$ 
      $\Rightarrow$ system now in state
      $\lambda_a\vec{s}(t)=\lambda_a(x(t),y(t),z(t))^T$ with
      \begin{equation}
        \label{eq:z}
        z(t)= \cos^2\theta + \sin^2\theta \cos(\omega t).
      \end{equation}
\item \textbf{Measure:} Repeat measurement and record outcome 
       $\lambda_b=\pm 1$
       $\Rightarrow$ system now in state $\lambda_b \vec{s}_0$.
\item \textbf{Repeat:} steps (i)--(iii) $N_e$ times.
\end{enumerate}
Let $N_{a,b}$ be the number of times the the initial measurement produced 
outcome $\lambda_a$ and the final measurement produced outcome $\lambda_b$.
Then there are four possible combinations of outcomes $N_{0,0}$, $N_{0,1}$,
$N_{1,0}$ and $N_{1,1}$, which must add to the total number of experiments
$N_e$.  The number of experiments that started in the state $\ket{0}$ [or
$\vec{s}_0=(0,0,1)^T$] is $N_e'=N_{0,0}+N_{0,1}$, and the fraction of 
experiments for which the second measurement yields $\lambda_0$ conditioned 
on the system starting in the state $\ket{0}$ is $N_{0,0}/N_e'$.  Thus the 
ensemble average of $\op{A}=\sz$ at time $t$, assuming we started in the 
state $\ket{0}$, is 
\begin{equation}
 \lambda_0 \frac{N_{0,0}}{N_e'} + \lambda_1 \frac{N_{0,1}}{N_e'}
 = \frac{\lambda_0 N_{0,0} + \lambda_1 N_{0,1}}{N_{0,0}+N_{0,1}},
\end{equation}
and for $N_e \to\infty$ this relative frequency should approach the
true expectation value of $z(t)$ assuming $z(0)=1$.
Thus, we can in principle determine $z(t)$ to arbitrary accuracy by
choosing $N_e$ large enough.  By repeating the experiments for
different evolution times $t_k$, e.g., using a stroboscopic mapping
approach (see Fig.~\ref{fig:stobo}), we can determine $z(t)$ as a
function of $t$, from which we can extract $\omega=\omega_{\vec{f}}$ 
and $\theta=\theta_{\vec{f}}$ as discussed.  For instance, noting that 
$z(t)=2p_0(t)-1$, Fig.~\ref{fig:leakage2} shows that the 1st order 
Fourier peak $F(\omega)$ of $z(t)$ in this example is $F(\omega)\approx
2\times 0.2468=0.4935$ for $\omega \approx 2$ and thus comparison
with Eq.~(\ref{eq:z}) shows that $\sin^2\theta = 2\Re[F(\omega)]$, or
equivalently, $\theta \approx 1.4566$.  These values are reasonably 
good estimates for the real values $\omega=2.0086$ and $\theta=-1.4780$,
except for the sign of $\theta$, or the orientation of the rotation
axis, which we cannot determine from the given data, as these rotations
have identical projections onto the $z$-axis.  This is reflected in
Eq.~(\ref{eq:z}) by the fact that both coefficients $\cos^2\theta$
and $\sin^2\theta$ contain squares.  In principle, we can determine
the parameters $\omega$ and $\theta$ to arbitrary accuracy using this
approach based on regular sampling and Fourier
analysis~\cite{PRA71n062312}, although the total number of experiments
can become prohibitively large.  An alternative that merits further 
investigation is the use of adaptive sampling techniques to reduce the 
total number of experiments necessary.

\subsection{Relative angles between rotation axes}

Having determined the parameters $\omega_{\vec{f}}$ and
$\theta_{\vec{f}}$ for a particular control setting $\vec{f}$ and chosen
a suitable reference Hamiltonian $\op{H}_{\rm ref}$ with $\phi_{\rm ref}=0$, 
to complete the characterization of
$\op{H}_{\vec{f}}\neq\op{H}_{\rm ref}$, we must determine the horizontal
angle $\phi_{\vec{f}}$.  The reference Hamiltonian $\op{H}_{\rm ref}$
must not commute with the measured observable $\op{A}$, or equivalently
the angle $\theta_{\rm ref}$ between the rotation axis $\vec{d}_{\rm
ref}$ and the $\vec{z}$-axis must be nonzero.  Ideally $\vec{d}_{\rm
ref}$ should be as close to orthogonal to the $\vec{z}$-axis as
possible.  Assuming $\theta_{\rm ref} \in (\pi/4,\pi/2]$, and
$\omega_{\vec{f}}$ and $\theta_{\vec{f}}$ are known, we can determine
the angles $\phi_{\vec{f}}$ by performing the following experiments%
~\cite{PRA69n050306}:
\begin{enumerate}
\item \textbf{Initialize:} Measure and record outcome $\lambda_a=\pm 1$ 
      $\Rightarrow$ system in state $\lambda_a\vec{s}_0=\lambda_a(0,0,1)^T$.
\item \textbf{Prepare:}
      Rotate around reference axis $\vec{d}_{\rm ref}$ by angle
      $\alpha_0=\arccos(\frac{1-x}{1+x})$, $x=-\cos(2\theta_{\rm ref})$ 
      $\Rightarrow$ system in new state 
      $\lambda_a\vec{s}_1=\lambda_a(\cos\beta,\sin\beta,0)^T$ with
      $\beta=\arctan(-\sqrt{2x})/\cos\theta_{\rm ref}$.
\item \textbf{Evolve:} Let the system evolve for time $t$ under the 
      control settings $\vec{f}$ 
      $\Rightarrow$ system in new state
      $\lambda_a \vec{s}(t) = \lambda_a (x(t),y(t),z(t))^T$ with
      \begin{subequations} 
      \label{eq:z2}
      \begin{align}
        z(t) &= c[1-\cos(\omega_{\vec{f}}t)]+d\sin(\omega_{\vec{f}}t)\\
        c    &= \sin\theta_{\vec{f}} \cos\theta_{\vec{f}} \cos(\phi_{\vec{f}}-\beta)\\
        d    &= \sin\theta_{\vec{f}}                      \sin(\phi_{\vec{f}}-\beta). 
      \end{align}
      \end{subequations}
\item \textbf{Measure:} Repeat measurement and record outcome
      $\lambda_b=\pm 1$
      $\Rightarrow$ system in new state $\lambda_b\vec{s}_0$.
\item \textbf{Repeat:} steps (i)--(iv) $N_e$ times.
\end{enumerate}
As before, if $N_{a,b}$ is the number of times the initial measurement 
produced outcome $\lambda_a$ and the final measurement produced outcome 
$\lambda_b$ then we have 
\begin{equation*}
 z(t)\approx \frac{\lambda_0 N_{0,0}+\lambda_1 N_{0,1}}{N_{0,0}+N_{0,1}} 
\end{equation*}
for sufficiently large $N_e$, thus allowing us to determine $z(t)$
experimentally.  Since $\theta_{\vec{f}}$, $\omega_{\vec{f}}$ and $\beta$ 
are known from the $\theta,\omega$ characterization step, Eq.~(\ref{eq:z2}) 
allows us to determine $\phi_{\vec{f}}$ via the coefficients $c$ and $d$, 
which can be determined in principle either through curve fitting or by 
taking the Fourier transform of $z(t)$ and noting that $c$ and $d$ 
correspond to the real and imaginary part of the 0th and 1st order
Fourier peaks.

\begin{figure}
\center{\includegraphics{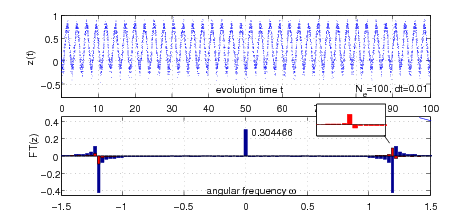}}
\caption{Determination of relative horizontal angle between two 
Hamiltonians.  Measured projection $z(t)$ of $\vec{s}_1(t)$ onto the 
$z$-axis and corresponding Fourier spectrum.  Blue/red bars indicate 
the cosine/sine coefficients, i.e., real/imaginary parts of the Fourier 
transform, respectively.}
\label{fig:phi}
\end{figure}

As a specific example, assume we have two Hamiltonians $\op{H}_1$ and 
$\op{H}_2$ and we have already determined $\omega_1=1$, 
$\theta_1=\frac{\pi}{4}$, and $\omega_2=1.2$, $\theta_2=\frac{\pi}{6}$.  
We choose $\op{H}_1$ as reference Hamiltonian and note that a rotation 
about $\op{H}_1$ by $\alpha=\pi$ maps the state $\vec{s}_0=(0,0,1)^T$ 
to $\vec{s}_1=(1,0,0)^T$.  We then let state $\vec{s}_1$ evolve for 
different length of time $t$ under Hamiltonian $\op{H}_2$, and determine 
the projection $z(t)$.  This will produce a measured $z$-trace as shown in 
Fig.~\ref{fig:phi}.  Eq.~(\ref{eq:z2}) shows that in theory the height 
of the real part of the 0th and 1st order Fourier peaks should be the 
same except for the sign, but the graph clearly shows that this is not 
the case for our noisy (simulated) experimental data.  In fact, the 1st
order peaks around $\pm \omega_{\vec{f}}$ are significantly broadened, 
while the $0$-peak is very narrow.  This suggests that we are likely to 
obtain the best estimate for $\phi$ by setting $c=\mbox{\rm FT}[z](0)$ 
and $\phi_{\vec{f}}=\arccos[c/(\sin\theta_{\vec{f}}\cos\theta_{\vec{f}})]
+\beta$, where $\beta=0$ in our case.  Indeed, this yields an estimate of 
$\phi_{\vec{f}}=0.2518 \pi$, which is very close to actual value $\pi/4$ 
used in the simulation.  Note that using the 1st order Fourier peaks 
gives far less accurate results due to the significant broadening of
the peaks.  This can be reduced using phase matching conditions and 
other tricks~\cite{PRA71n062312}, but in this case simply using only 
the 0-peak is sufficient to get a good estimate.

\subsection{Functional dependence on control settings}

In some settings it is sufficient to characterize the Hamiltonians
$\op{H}_{\vec{f}}$ for a discrete, finite set of controls
$\{\vec{f}^{(k)}\}$.  This would be the case for bang-bang-type control
schemes that require only switching between a finite set of control
Hamiltonians.  In other cases, in particular when continuously varying
fields are to be applied, we would also like to know the functional
dependence of the Hamiltonian on the control fields, e.g., whether we 
can assume a control-linear model as in (\ref{eq:Hf}), or if there
are non-linear or crosstalk effects, etc.  Although characterizing the 
control dependence of the Hamiltonian is a non-trivial problem, given 
sufficiently many $\op{H}_{\vec{f}}$ data points, it is possible to 
test the quality of fit of a particular model using statistical means.  
For instance, suppose we have characterized the Hamiltonians
$\op{H}_{\vec{f}}$ for $\vec{f}\in\{\vec{f}^{(k)}\}$.  Let $\vec{d}_k$
be the corresponding vectors in $\RR^3$ as defined above, and let
$f_m^{(k)}$ be the $m$th component of the control vector
$\vec{f}^{(k)}$.  We can find the best-fitting linear model by finding
vectors $\tilde{\vec{d}}_m \in \RR^3$, $m=0,\ldots,M$, where $M$ is 
the number of independent control variables, that minimize the norm of
the residuals $\Delta=\sqrt{\sum_k \norm{\vec{r}_k}_2^2}$ where
\begin{equation}
   \vec{r}_k = \tilde{\vec{d}}_0 + \sum_{m=1}^M f_m^{(k)} 
                                   \tilde{\vec{d}}_m - \vec{d}_k.
\end{equation}
The norm of the residuals $\Delta$ then gives an indication of the 
quality of fit of a linear dependence model, and allows us to compare 
the quality of fit for different models.

\begin{figure}
\center{\includegraphics{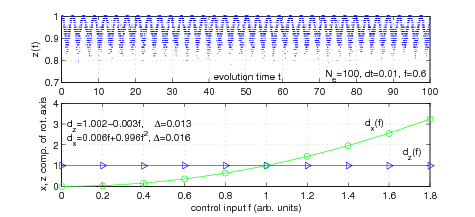}}
\caption{Determination of functional dependence of $\op{H}_{\vec{f}}$
on control input $\vec{f}$.  The bottom graph shows the estimated values
for $d_x^{(k)}$ and $d_z^{(k)}$ (circles and triangles, respectively) as
well as the actual functional dependence in the model (solid lines). 
The top graph shows a typical $z(t)$ trace obtained by averaging over 
$N_e=100$ simulated experiments for each time $t_k=0.01 k$.  The effects 
of finite sampling and projection noise are clearly visible for this
trace.  Nonetheless, the predicted functional dependence of $\op{H}_{f}$ 
on the control input $f$ is in good agreement with the actual dependence 
in the model.}
\label{fig:fdepend}
\end{figure}

As an example, consider a Hamiltonian $\op{H}_{\vec{f}}$ depending on
a one-dimensional control input $\vec{f}=f$.  To characterize the 
functional dependence of $\op{H}_f$ on $f$, we choose a finite set of
control settings $f_k=0.2(k-1)$ for $k=1,\ldots,10$.  We characterize
first the rotation frequencies $\omega_k$ and declination angles 
$\theta_k$ for each $f_k$, and then the relative angles $\phi_k$ as
discussed above.  From these, we calculate the components $d_x^{(k)}$, 
$d_y^{(k)}$ and $d_z^{(k)}$ of the Hamiltonian according to 
Eq.~(\ref{eq:H2}), and plot
them versus $f$ as shown in Fig.~\ref{fig:fdepend}.  Specifically, in 
our case we find $\phi_k\approx 0$ and hence $d_y\equiv 0$, $d_z$ 
appears to be constant and $d_x$ exhibits a quadratic dependence on 
the control $f$.  A least square fit of the data using a linear and
quadratic fitting function, respectively, yields
$d_z(f)=1.002-0.003\,f$ and $d_x(t)=0.006f+0.996f^2$, which is in 
very good agreement with the actual functional dependence in the model
$d_z(f)=1$ and $d_x(f)=f^2$ despite the fact that the simulated data
appears very noisy and the $\theta$ and $\omega$ estimates did not
employ tuning such as phase matching to further improve the accuracy.

\subsection{Characterization in the presence of decoherence}

The protocols for Hamiltonian identification above are sufficient if 
dissipative effects are small on the timescales for coherent control,
i.e., if the system and environment coupling is much weaker than the
system-control interaction.  If the coupling between the system and 
its environment is stronger then dissipative effects must be directly
incorporated into the characterization protocols.
 
Assuming Markovian decoherence of the form~(\ref{eq:diss}), it is easy
to show that for $N=2$, we have $\op{A}_k=\op{U}\s_k\op{U}^\dagger$, 
where $\s_k$ are the elementary relaxation operators 
$\{\s_-,\s_+,\frac{1}{2}\sz\}$ defined in terms of the Pauli matrices
$\s_{\pm}=\frac{1}{2}(\sx\mp\rmi\sy)$, and $\op{U}$ is some unitary 
operator in $\SU(2)$ that defines a preferred decoherence basis.  The 
operator $\sz$ is usually associated with pure decoherence (pure 
dephasing) while the raising and lowering operators $\s_{\pm}$ are 
associated with population relaxation from $\ket{1}$ to $\ket{0}$ and 
vice versa.  Very often it is furthermore assumed that $\op{U}=\op{I}$
is the identity, i.e., that the preferred basis for relaxation is the
same as the measurement basis.  Although this need not be the case, it
is often a good approximation.  The main effect of relaxation processes 
is to dampen the observed coherent oscillations $z(t)$ that result from 
the Hamiltonian dynamics, and try to force the system `asymptotically' 
into a `steady' state.  This effect manifests itself in Lorentzian 
broadening of the peaks in the oscillation spectrum, which can be used 
for characterization purposes.

\begin{figure}
\center{\includegraphics{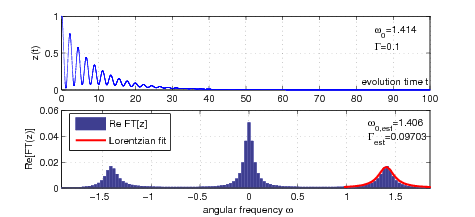}}
\caption{Characterization in the presence of decoherence.  The observed
damped oscillations due to pure dephasing (top) result in Lorentzian
broadening of the Fourier peaks, which can be exploited to obtain (good)
estimates for the rotation frequency $\omega_0$ and decoherence rate 
$\Gamma$.}
\label{fig:decoherence}
\end{figure}

For example, Fig.~\ref{fig:decoherence} shows significantly damped
oscillations as a result of relaxation.  The fact that the oscillations 
decay to $z_\infty=0$ suggests either pure dephasing, or a symmetric 
relaxation process with $\Gamma_{+}=\Gamma_{-}$, i.e., equal probability 
of relaxation from state $\ket{0}$ to $\ket{1}$ and vice versa, or a
combination of both.  Since symmetric relaxation and pure dephasing 
have very similar signatures, it can be difficult to distinguish these
processes from the oscillation trace alone.  However, if there exists 
a control $\vec{f}$ such that $\op{H}_{\vec{f}}\propto \sz$, which
corresponds to a rotation about the $\vec{z}$ (or measurement) axis,
then we can in principle distinguish the two processes by initializing
the system repeatedly in either state $\ket{0}$ or $\ket{1}$ and letting
the system evolve under $\op{H}_{\vec{f}}$ for various periods of time
before measuring again.  If the decoherence process is pure dephasing
then the system will remain in the initial state, while we expect to
observe random jumps (with equal probability) for symmetric relaxation.
Assuming this preliminary characterization step suggests that the damping
in the figure is due to pure dephasing, we can then extract information
about the rotation frequency $\omega_0$ and dephasing rate $\Gamma$ by 
fitting a Lorentzian envelope function
\begin{equation}
  L_{\omega_0,\Gamma}(\omega) = \frac{\Gamma}{(\omega-\omega_0)^2+\Gamma^2} 
\end{equation}
to the first-order peak in the Fourier spectrum as shown in the figure.  
In our example, using a simple least-squares minimization yields
estimated values for both $\omega_0$ and $\Gamma$ that are close to the
actual values used in the simulated experiment.  This approach can be
generalized for more complicated relaxation processes~\cite{PRA73n062333}.
Alternatively, if we can characterize the Hamiltonian dynamics sufficiently 
to be able to initialize the system in non-measurement basis states and 
simulate readout in different bases, then generalized Lindblad operators 
can be estimated using repeated process tomography~\cite{PRA67n042322}.

\section{Discussion of generalizations and conclusions}
\label{sec:conclusion}

We have described some basic tools and techniques for characterizing 
the extent to which the dynamics of a certain system is confined to a
low-dimensional subspace of a potentially large Hilbert space, as well
as for identifying both Hamiltonian and decoherence parameters  
experimentally, with emphasis on protocols that are realistic even in 
the presence of limited direct readout and control capabilities.  

Although the explicit schemes presented focused on qubit-like systems, 
the same basic ideas can be applied to higher-dimensional systems.  
However, the number of Hamiltonian and decoherence parameters to be
determined for higher-dimensional system makes finding explicit
protocols for complete characterization quite challenging.  The qubit 
characterization protocols show that identifying the relative angles 
between several control Hamiltonians requires a two-stage process and
more complicated two-step experiments (initialization in non-measurement 
basis state followed by controlled evolution) than identifying the 
rotation frequency and declination angle for a single Hamiltonian.  
As the dimension of the system increases more parameters are necessary 
to fully characterize the Hamiltonians, leading to more complex 
multi-stage characterization protocols.

Another problem is that the number of experiments required to implement
such schemes may become prohibitively large for parameter estimation
based on spectroscopic analysis.  One of the most promising strategies
to avoid this problem is the development of efficient adaptive
characterization schemes---as opposed to protocols based on regular
sampling--- to minimize the number of measurements / experiments
required.  For systems composed of separate smaller units such as
multi-qubit systems, boot-strapping approaches to characterization may
also provide a fruitful alternative.  For example, assuming we have
separately characterized individual qubits using qubit identification
protocols and therefore have full local control, we can reduce the
problem of two-qubit identification in principle to identifying the
interaction Hamiltonian.  For two-qubit systems with local control and
a fully non-local interaction Hamiltonian this can be achieved using 
entanglement mapping or concurrence 
spectroscopy~\cite{devitt:052317, JPA39p14649}.

\ack
SGS acknowledges funding from an \mbox{EPSRC} Advanced Research
Fellowship and support from Hitachi and the \mbox{EPSRC}
\mbox{QIP IRC}.  DKLO is supported by a SUPA fellowship.  SJD
acknowledges support from MEXT.

\providecommand{\newblock}{}

\end{document}